# Data harvesting vs data farming: A study of the importance of variation vs sample size in deep learning-based auto-segmentation for breast cancer patients

Emma Skarsø Buhl[1,2], Else Maae[3], Louise Wichmann Matthiessen[4], Mette Holck Nielsen[5], Maja Maraldo[6], Mette Møller[7], Stine Elleberg[1], Sami Aziz-Jowad Al-Rawi[8], Birgitte Vrou Offersen[9], and Stine Sofia Korreman[1,2]

[1]Danish Center for Particle Therapy, Aarhus University Hospital, Aarhus, Denmark
[2]Department of Clinical Medicine, Aarhus University, Aarhus, Denmark
[3]Department of Oncology, Vejle Hospital, University Hospital of Southern Denmark, Denmark
[4]Department of Oncology, Herlev and Gentofte Hospital, Herlev, Denmark
[5]Department of Oncology, Odense University Hospital, Denmark
[6]Department of Clinical Oncology, Rigshospitalet, Copenhagen University Hospital, Denmark
[7]Department of Oncology, Aalborg University Hospital, Aalborg, Denmark
[8]Department of Clinical Oncology and Palliative Care, Zealand University Hospital, Næstved, Denmark
[9]Department of Experimental Clinical Medicine, Aarhus University Hospital, Aarhus, Denmark

**Abstract** Large interobserver variations (IOV) are expected when delineating structures on a CT scan, especially small structures with low visibility, like the clinical target volume internal mammary lymph node chain (IMN). A solution to avoiding IOV is deep learning-based auto-segmentation (DL). However, to train a robust and well performing model, a large and high-quality data set is needed. A large data set can become available if using delineations from the daily clinical practice, however these can be prone to large IOV. A data set created specifically for the purpose of training a DL model allows for obtaining a lower IOV, however this also entails a small cohort, as it is time-consuming to create. The aim of this study was to investigate the difference in output, when training a model in three different scenarios: a large clinical delineated data set (with 700/78 patients for training/testing, from the Danish Breast Cancer Group (DBCG) RT Nation Study), a clinical but curated dataset (with 328/36 patients for training/testing, from the DBCG RT Nation Study) and a smaller, but dedicated data set created by delineation experts (with 123/14 patients for training/testing, consensus delineations created by delineation experts). The model performance was estimated based on the performance metrics dice similarity coefficient (DSC), Hausdorff 95$^{th}$ percentile (HD95) and mean surface distance (MSD). Models were tested in test sets from their own cohort, and afterwards also compared in the dedicated data test set. The difference between model output was finally estimated by measuring the mean width and cranial caudal length of the model output for the models. When testing the model output between the clinical models and the dedicated models in their own test set, the two clinical models had a poorer performance, than the dedicated models, but not all metrics showed statistically significance. When testing the models in the dedicated data, the dedicated model showed a slightly better performance, along with fewer segmentation outliers. As a way of taking advantage of the strength from both types of data set, it could be an option to use a large clinical data set as a baseline model, and then finetune with smaller sized cohorts with dedicated delineations.

## 1 Introduction

Yearly more than 4700 women are diagnosed with breast cancer in Denmark[1]. Most patients are operated and more than 80% of these are treated with post-operative adjuvant radiotherapy[2]. Node positive breast cancer patients are treated with internal mammary lymph node (IMN) irradiation to improve overall survival[3,4]. Constraints to organs at risk (OAR), such as the heart, lung and contralateral breast, often call for a clinical compromise during treatment planning. The radiation treatment plans are optimized based upon manual structure delineations on CT scans. The IMN target is difficult to see on CT scans[5], which may introduce interobserver variation (IOV), despite strict national and international delineation guidelines[5,6]. Recently, DL has been studied as a way of mitigating IOV. For OAR, DL has been shown to work well [7], but for target structures there are still challenges. In order to achieve a good performance with DL models, large data sets of high quality are generally needed. But, achieving both high quantity and high quality in these data sets is a challenge, as delineation is time-consuming.

In the radiotherapy community, the data sets for DL models are often achieved through data farming - a single expert or a small group delineates a dedicated data set, to minimize the IOV. This is only feasible for small data sets due to the limit on clinicians' time, and thus hampers the size of the cohort. This further decreases the generalizability of the model, as the limited size training data set lacks diversity. One solution to obtaining large data sets could be to use the clinical delineations retrospectively. This will not only allow for larger data sets, but also include a variation of patient anatomy, setup scenarios and CT scanners for increased generalizability. Clinical delineations reflect daily high workload and a larger variation in delineation quality is therefore expected compared to dedicated data sets created by a single or group of experts strictly following the same delineation guidelines. The Danish Breast Cancer Group (DBCG) has established the DBCG RT Nation database, containing CT scans and structure sets from more than 7500 Danish high-risk breast cancer patients, who received radiotherapy in 2008-2016[8]. In addition, for a small subset of the DBCG RT Nation data set dedicated edited delineation of target structures has been generated by the DBCG following a target delineation consensus workshop.

The primary aim of this study was to investigate the difference in model performance when training a DL based model for the IMN target structure on clinical delineations vs dedicated delineations.

## 2 Materials and Methods

The study included left-sided breast cancer patients, treated in 2015-2016 from the DBCG RT Nation study, as this was the period after implementation of national delineation guidelines[5,6]. In total 778 patients were included (data_CLIN). The patients were treated with adjuvant loco-regional radiotherapy at the seven radiotherapy centers in Denmark, generating a variance in scanner and set-up protocols in the data. Depending on tumor location, it is clinical practice to delineate the IMN to intercostal room (IC) 3 or IC 4.

The full CT scans and IMN delineations were used as input in a 3D full-resolution nnUNet [9], and the data set was randomly split into a training set (90%) and test set (10%). Five-fold cross validation was used with default parameters and the number of epochs was varied depending on cohort size, to avoid overfitting. Before training of the two models, the DICOM files were converted into NIfTI files with the Python package dcmrtstruct2nii [10]. Three different models were trained. The first model was trained on the full cohort of 778 clinical IMN delineations (model_CLIN), with no data curation. The model was trained with 800 epochs. For the second model, the data was crudely curated to only encompass IMN delineations extending to IC 3, to lower the variance in the data set occurring due to differences in inclusion of IC 3 and IC 4 (data_CURATED_CLIN). The model was trained with 600 epochs (model_CURATED_CLIN). The third model was trained on a smaller, but dedicated data set (data_DED). The data_DED was created from the data_CLIN. Of the 778 left-sided breast cancer patients, 143 were used to create the data_DED. The patients were chosen to be equally distributed from the seven radiotherapy centers in Denmark and equally distributed between lumpectomy or mastectomy. A national target delineation workshop was held in October 2022 by the DBCG and the Danish Comprehensive Cancer Center, where 21 participants from all RT centres in Denmark discussed the national delineation guidelines [5,6]. Following the workshop days, the participants were asked to correct the 143 clinical delineations (6-7 each), so they followed the national delineation guidelines. The model_DED was trained with 300 epochs. Because the dedicated data set was relatively small, as an extra robustness test, the training and test set split was performed arbitrarily additionally two times with the data_DED. This allowed for an extra two versions of model_DED (model_DED_second_split and model_DED_third_split) and two extra dedicated test sets, to test the performance of model_CLIN and model_CURATED_CLIN. To estimate and compare the IOV in the five training cohorts, the mean width and cranial caudal length were calculated for all delineations in the training cohort. The width was calculated by measuring the longest distance between each coordinate set, and subsequently the mean width was calculated by averaging over all slices. The cranial caudal distance was measured from the most cranial slice to the most caudal slice. The measurements were conducted on the NIfTI masks.

The performance of the five models was first evaluated in their own test-set with the performance metrics: the dice similarity coefficient (DSC), the Hausdorff distance 95th percentile (HD95) and the mean surface distance (MSD). The statistical significance was tested with the Mann-Whitney U-test, assuming statistical significance level at $p<0.05$. As the variation in model training data might be reflected in the testing data as well, the model_CLIN and model_CURATED_CLIN was tested in the three different data_DED test sets, to estimate the performance of all three in a low IOV data set. In addition, the mean width and the cranial caudal length of the three model_DED, model_CLIN and model_CURATED_CLIN segmentations in the model_DED test sets were measured and compared. The statistical significance was tested with the Wilcoxon signed-rank test, assuming statistical significance level at $p<0.05$.

|  | DSC (mean) range | HD95 [mm] (Mean) range | MSD [mm] (mean) range |
|---|---|---|---|
| Model_CLIN (own test set, n = 78) | 0.65 *,+, # (0.46 – 0.80) | 11 *,# (2.30 – 125.6) | 2.55 *,+,# (0.59 – 19.8) |
| Model_CLIN (dedicated test set first split, n = 14) | 0.71 (0.57– 0.84) | 6.54 (1.95–25.8) | 1.37 (0.42 – 3.49) |
| Model_CLIN (dedicated test set second split, n = 14) | 0.66 (0.54 – 0.84) | 10.2 (1.95 – 32.0) | 2.14 (0.42–7.97) |
| Model_CLIN (dedicated test set third split, n = 14) | 0.69 ¤ (0.58 – 0.81) | 8.69 (2.34– 28.4) | 1.65 ¤ (0.48–6.10) |
| Model_CURATED_CLIN (own test set, n = 36) | 0.66 # (0.36 – 0.80) | 11.8 # (2.16-43.2) | 3.00 *,# (0.54 – 11.8) |
| Model_CURATED_CLIN (dedicated test set first split, n = 14) | 0.69 (0.52 – 0.81) | 8.60 (2.62– 23.4) | 1.20 (0.58–2.33) |
| Model_CURATED_CLIN (dedicated test set second split, n = 14) | 0.64 ¤ (0.51 – 0.78) | 13.3 (3.00– 30.0) | 1.65 (0.39–6.16) |
| Model_CURATED_CLIN (dedicated test set third split, , n = 14) | 0.67 ¤ (0.53 – 0.77) | 9.39 ¤ (2.54– 33.5) | 1.05 (0.51–2.22) |
| Model_DED (dedicated test set first split, n = 14) | 0.71 (0.61-0.86) | 7.58 (1.95-15.4) | 1.41 (0.30-3.31) |
| Model_DED_second_split (dedicated test set second split, n = 14) | 0.71 (0.57-0.83) | 5.70 (1.95-14.0) | 1.20 (0.54-2.44) |
| Model_DED_third_split (dedicated test set third split, n = 14) | 0.74 (0.62-0.85) | 5.30 (2.00-18.4) | 0.96 (0.45-2.64) |

Table 1. The performance of model_CLIN, model_CURATED_CLIN and model_DED in their respective test set and in the three dedicated test set splits (n=14). The performance metrics are the dice similarity coefficient (DSC), the Hausdorff 95th percentile (HD95) and the mean surface distance (MSD). Statistical significance between clinical models tested in their own test and model_DED is denoted with * for the first split, + for the second split and # for the third split. Statistical significance between clinical models tested in the dedicated test sets and model_DED tested in the corresponding split is denoted ¤. The third test is used as example in both figure 1 and 2.

## 3 Results

A total of 778 delineations were available from 2015-2016, leaving 700 for training the model_CLIN and 78 for testing the model_CLIN. During curation of the IMN delineations a total of 442 delineations were excluded, leaving 328 for training the model_CLIN_CURATED and 36 for testing the model_CLIN_CURATED. In the data_DED six delineations were excluded, due to the IMN extending to IC 2 or IC 4, leaving 123 for training the model_DED and 14 for testing the model_DED. In the training cohort, the data_CLIN had a mean width = 14.4 mm and a mean cranial caudal length of 108.2mm, the data_CURATED_CLIN had a mean width = 13.5mm and a mean cranial caudal length of 102.1mm and the data_DED (first/second/third) with a mean width = 14.5 mm/14.4mm/14.4mm and mean cranial caudal length of 107.8mm/108.1mm/108.0mm, figure 1.

Statistically significant difference was found for all except the difference in cranial caudal length between data_CLIN and data_DED and cranial caudal length and width between data_CLIN and data_DED_second_split and data_DED_third_split.

In their respective test sets, the two clinical models generally had poorer performance (mean DSC 0.65-0.66, HD95 11-11.8, MSD 2.55-3.00) than the three model_DED (mean DSC 0.71-0.74, HD95 5.3-7.58 MSD 0.96-1.41), however not all metrics showed statistical significance, table 1.

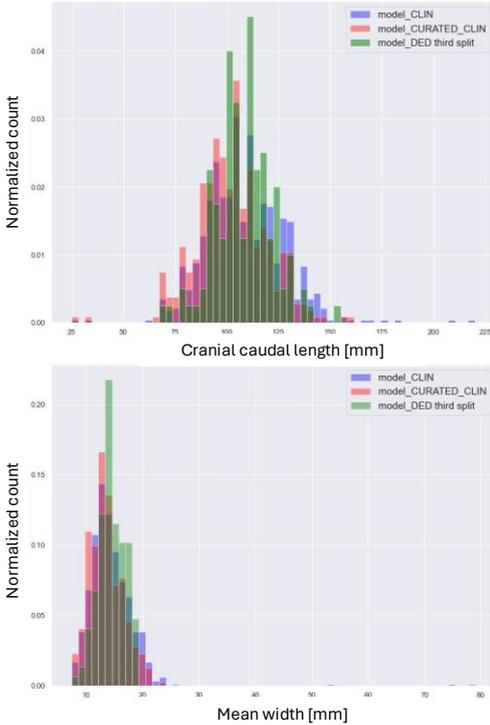

**Figure 1:** (top) The cranial caudal length of the delineations in the training cohort of the model_CLIN(blue), model_CURATED_CLIN (orange) and the third split of model_DED (green). (bottom) the mean width of the delineations in the training cohorts The counts are normalized.

The difference between the performance of the two clinical models tested in the three dedicated test sets and the model_DED varied with the different training/test splits, table 1. A large range in both HD95 and MSD was seen in the two clinical models (MSD range 0.42-7.97 and 0.39-6.16) when tested on the dedicated test sets, along with an overall better performance of the three model_DED (MSD range 0.54-2.44), but not all were statistically significant, table 1.

The mean width of the CTVn_IMN with the model_CLIN in the third dedicated test set was 12.9mm and the mean cranial caudal length was 115.7mm and the mean width with the model_CURATED_CLIN was 12.3mm and the cranial caudal length was 104.7mm. The mean width of the CTVn_IMN with the model_DED_third_split in the third dedicated test set split was 137.7mm and mean cranial caudal length was 112.1mm. The difference in cranial caudal length was found statistically significant between model_DED_third_split and model_CURATED_CLIN and the width was found statistically significant between both clinical models and the model_DED_third_split, figure 2 and table 2.

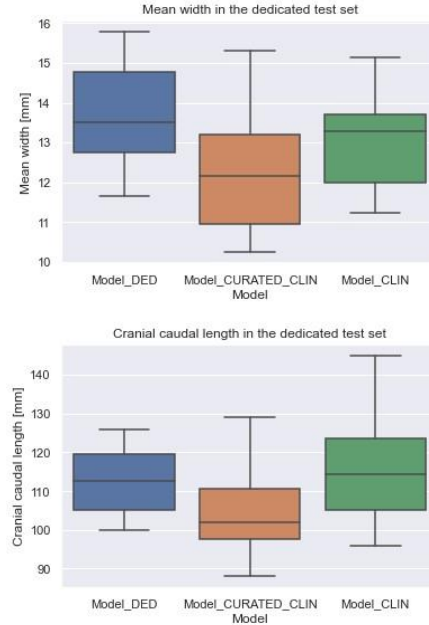

**Figure 2:** (top) The mean width in model output of the third split with model_DED (blue), model_CURATED_CLIN (orange) and model_CLIN (green) in the dedicated test - set. (Bottom) the cranial caudal length of the segmentations.

|  | Cranial caudal length [mm] (mean) range | Width [mm] (mean) range |
|---|---|---|
| Model_CLIN (dedicated test set first split, n = 14) | 119.4 (102.0-150.0) | 12.9 (11.0-14.8) |
| Model_CLIN (dedicated test set second split, n = 14) | 113.3 (97.5-132.0) | 13.3 * (11.0-16.0) |
| Model_CLIN (dedicated test set third split, n = 14) | 115.7 (96.0-145.0) | 12.9 * (11.2-15.1) |
| Model_CURATED_CLIN (dedicated test set first split, n = 14) | 110.0 (86.3-144.0) | 12.4 * (10.8-14.7) |
| Model_CURATED_CLIN (dedicated test set second split, n = 14) | 101.3 * (85.0-129.0) | 13.3 * (11.1-17.0) |
| Model_CURATED_CLIN (dedicated test set third split, , n = 14) | 104.7 * (88.0-129.0) | 12.3 * (10.2-15.3) |
| Model_DED (dedicated test set first split, n = 14) | 115.1 (94.0-130.0) | 13.4 (11.6-16.1) |
| Model_DED_second_split (dedicated test set second split, n = 14) | 112.5 (92.5-136.0) | 14.2 (12.2-16.7) |
| Model_DED_third_split (dedicated test set third split, n = 14) | 112.1 (100.0-126.0) | 13.7 (11.6-15.8) |

**Table 2.** The mean width and cranial caudal length with the model_CLIN, model_CURATED_CLIN and the three model_DED tested in the three dedicated test sets. Statistically significant difference between the clinical models and the respective dedicated model is shown with *.

## 4 Discussion

In this study, we have investigated the difference in performance when using a clinical cohort, a curated clinical cohort, and a dedicated cohort to train DL-based auto-segmentation models for the left-sided CTVn_IMN. The differences among the three model predictions were minor and dependent on test set choice. However, for geometrical shape metrics, statistically significant differences were seen between the clinical models and the three versions of model_DED. The optimal balance between high-quality delineations and a large cohort is difficult, and both factors contribute to the performance of a DL model. Normal practice in the radiotherapy community is data farming, by creation of a dedicated data set, which limits the cohort size. Studies have investigated how low it is reasonable to go in size of data set without losing the precision of the DL model. In Henderson et al, it was shown that for OAR in head and neck cancer increasing the data set beyond 250 scans gave very little improvement in model performance[11]. However, in a recent study from our group, we showed that for heart segmentation, smaller training cohorts increased the risk of outlier segmentations, when testing it in a set with larger variety in anatomy[12].

In this present study, there is a difference in model performance metrics among the three models, however there is not a consistent statistically significant difference between the models. It is worth noting that the test sets with dedicated delineations were small (n=14) and therefore does not represent a large variety of patient anatomies. Two additionally splits of the train/test set was performed with data_DED as an extra robustness test. The three different data_DED test sets led to different results of performance for model_CLIN and model_CURATED_CLIN, showcasing the impact of the specific patients included in the test set for such a small number of patients[12]. It would be expected that a larger number of patients in the test set would give a better estimate of the robustness of the models. However, despite the small test sets, there is still evidence of generally poorer performance and larger spread in the range of performance metrics in the clinical models than in the dedicated models, table 1. As a way of taking advantage of the strength from both types of data set, it could be an option to use a large clinical data set as a baseline model, which can then be finetuned with smaller sized cohorts with dedicated delineations. This would give both the variety available for generalizability from the clinic, and the low IOV and high precision delineations from dedicated data generation. If this approach is proven feasible, it might also be a way to tackle the problems arising when the delineation guidelines are updated and there will be a need for an update of the DL model. The dedicated delineations were created by delineation experts from seven institutions and based on editing existing delineations used in the clinic. The choice of correcting clinical delineations instead of delineation from scratch aimed to lower the delineation burden on busy clinicians. However, presenting delineators with an existing delineation imposes a risk of creating a bias as compared to delineating from scratch. The variety of expert delineators also introduce some IOV. Nevertheless, it is expected that the delineations created in the dedicated process, are as precise as possible on a national scale. Only a quantitative comparison has been conducted on these models. For a performance check, a qualitative investigation with expert assessments would be needed, as studies have shown, that the quantitative metrics not always correlate with clinical acceptance[13].

## 5 Conclusion

Training of a model on a clinical data set, showed on average a good performance, showcasing the benefits of using a large data set. Furthermore, a model trained on a large clinical data set performed better than a model trained on a curated clinical data set. However, the models trained on a dedicated data set showed a slightly improved performance, along with fewer segmentation outliers as compared to the models trained on clinical delineations. A combination of both models may be a way to utilize the best of both worlds.